\documentclass[%
reprint,
showpacs,
 amsmath,amssymb,
 aps,
 prl
]{revtex4-1}

\newcommand{\ban}{\begin{eqnarray*}}
\newcommand{\ean}{\end{eqnarray*}}

\newcommand{\bit}{\begin{itemize}}
\newcommand{\eit}{\end{itemize}}

\usepackage{graphicx}
\usepackage{dcolumn}
\usepackage{bm}

\begin{document}


\title{Electroweak Sudakov Corrections to New Physics Searches 
at the CERN LHC}

\author{Mauro Chiesa} \email{mauro.chiesa@pv.infn.it}
\author{Guido Montagna}%
 \email{guido.montagna@pv.infn.it}
\affiliation{%
 Dipartimento di Fisica, Universit\`a di Pavia, and \\INFN, 
Sezione di Pavia, Via A. Bassi 6, 27100 Pavia, Italy
}%
\author{Luca Barz\`e} \email{luca.barze@cern.ch}
\affiliation{CERN, PH-TH Department, Geneva, Switzerland
}%
\author{Mauro Moretti} \email{mauro.moretti@fe.infn.it}
\affiliation{Dipartimento di Fisica e Scienze della Terra, 
Universit\`a di Ferrara, and \\INFN, Sezione di Ferrara, Via Saragat 1, 
44100 Ferrara, Italy
}%
\author{Oreste Nicrosini}\email{oreste.nicrosini@pv.infn.it}
\author{Fulvio Piccinini}\email{fulvio.piccinini@pv.infn.it}
\affiliation{%
INFN, Sezione di Pavia, Via A. Bassi 6, 27100 Pavia, Italy
}%
\author{Francesco Tramontano}\email{francesco.tramontano@cern.ch}
\affiliation{%
Dipartimento di Scienze Fisiche, Universit\`a di Napoli ``Federico II", 
and \\INFN, Sezione di Napoli, via Cintia, 80126 Napoli, Italy
}%

\date{\today}

\begin{abstract}
We compute the one-loop electroweak Sudakov corrections to the production 
process $Z (\nu \bar{\nu}) + n$~jets, with $n = 1,2,3$, in $p p$ 
collisions at the LHC. It represents the main irreducible background 
to new physics searches at the energy frontier. The results are
obtained at the leading and next-to-leading logarithmic accuracy 
by implementing 
the general algorithm of Denner-Pozzorini in the 
event generator for multiparton processes {\tt ALPGEN}. 
For the standard selection cuts 
used by ATLAS and CMS collaborations, we 
show that the Sudakov corrections to the relevant observables can grow 
up to $ - 40$\% at $\sqrt{s} = 14$~TeV. 
We also include the contribution due to undetected real radiation of 
massive gauge bosons, to show to what extent the partial cancellation with 
the large negative virtual corrections takes place in realistic event 
selections.
\end{abstract}

\maketitle



Important searches for new phenomena beyond the Standard Model 
(SM) of particle physics at the 
proton-proton ($pp$) collider LHC at the CERN laboratory are based on the 
analysis of 
events with jets and missing transverse momentum ($\rlap\slash{\!p_T}$). 
Typical examples of such studies are searches for squarks and gluinos in 
all-hadronic reactions containing high-$p_T$ jets, missing transverse momentum 
and no electrons or muons, as predicted in many supersymmetric extensions 
of the SM. 
These final states can appear in a number of R-parity conserving models where 
squarks and gluinos can be produced in pairs and subsquently decay to standard 
strongly interacting particles plus neutralinos that escape 
detection, thus giving rise to a large amount of 
$\rlap\slash{\!p_T}$. Typically the event selections adopted require 
the leading 
jet $p_T$ larger than 130~GeV or the single  jets $p_T$'s larger than 50~GeV. 
Moreover, the signal region is defined by $m_{\rm eff} > 1000 $~GeV, 
where $m_{\rm eff} = \sum_i  {p_T}_i   + \rlap\slash{\!E_T}$, 
or $H_T > 500$~GeV and $| \vec{\rlap\slash{\!H_T}} | > 200$~GeV, 
where $H_T = \sum_i {p_T}_i$ and 
$\vec{\rlap\slash{\!H_T}} = 
- \sum_i \vec{p_t}_i$~\cite{Aad:2011ib,Collaboration:2011ida,cms:2012mfa}. 

The main SM backgrounds to the above mentioned signal(s) are given by the 
production of weak bosons accompanied by jets ($W/Z + n$~jets), 
pure QCD multiple jet events and $t \bar{t}$ production. Among these 
processes only $Z + n$~jets (in particular with $Z \to \nu \bar\nu$) 
constitutes an irreducible background, particularly relevant for final states 
with 2 and 3 jets. Because new physics signals could manifest themselves 
as a mild deviation with respect to  the large SM background, precise 
theoretical predictions for the processes under consideration are needed. 
Moreover, for these extreme regions, it is known that the observables 
are affected by large electroweak (EW)  Sudakov corrections. The aim of the 
present paper is to compute  the one-loop EW Sudakov corrections to the 
production process $Z (\nu \bar{\nu}) + n$~jets, with $n = 1,2,3$, in $p p$  
collisions at the LHC. 
It is worth noting that the experimental procedure for the 
irreducible background determination relies on data driven methods of 
measuring control samples of either $\gamma +$~jets, $Z(\to l^+ l^-) +$~jets 
and $W(\to l \nu_l) +$~jets. Therefore, the required theoretical information 
is the prediction of the ratios of cross sections for the above three 
processes. In the ratios the uncertainties related to QCD and PDF's 
largely cancel, while the electroweak 
corrections do not~\footnote{We thank G. Watt for pointing this out. 
This has been studied for the case of $V + 1$~jet 
in Ref.~\cite{Malik:2013kba} and we leave the analogous analysis 
for $V + $~multijets to a future study. }.

Before discussing the details of the calculation, let us summarize 
the available QCD/EW calculations for the processes 
$V=W,Z$ + jets. Exact NLO QCD corrections to $Z + 4$~jets and $W + 4$~jets, 
computed by means of the package {\tt BlackHat} and interfaced to the parton 
shower generator {\tt SHERPA} can be found 
in Refs.~\cite{Ita:2011wn,Bern:2012vx} 
and~\cite{Berger:2010zx}, respectively. 
Fixed-order (NLO) QCD predictions for the production of a vector boson in 
association with 5 jets at hadron colliders are presented in 
Ref.~\cite{Bern:2013gka}.
Leading and next-to-leading EW corrections to the processes 
$V= \gamma,Z,W$ + 1~jet, 
with on-shell $W,Z$ bosons, can be found in 
Refs.~\cite{Kuhn:2005gv, Kuhn:2004em, Kuhn:2007qc}, where  two-loop Sudakov 
corrections are also investigated. Very recently EW and QCD corrections 
to the same processes 
have been computed using the soft and collinear effective theory 
in~\cite{Becher:2013zua}. 
The exact NLO EW calculation for $V=W,Z$ + 1~jet, with on-shell $W,Z$ bosons 
can be found in Refs.~\cite{Kuhn:2005az,Kuhn:2007cv,Maina:2004rb}, and 
the same with $W, Z$ decays 
has been published in Refs.~\cite{Denner:2012ts,Denner:2011vu,Denner:2009gj}. 
Recently the exact NLO EW calculation for $Z (\nu \bar\nu) +$~2~jets, for the 
partonic subprocesses 
with one fermion current only 
(i.e. including only gluon-gluon ($g g$) contributions 
to 2 jets), has been completed and can be found 
in Ref.~\cite{Actis:2012qn}. 
No EW corrections for $Z + 2/3$~jets production including all partonic 
subprocesses are available at the moment. 

For energy scales well above the EW scale, EW radiative corrections are 
dominated by double and single logarithmic contributions 
(DL and SL, respectively) whose argument involves the ratio of the energy 
scale to the mass of the weak bosons. 
These logs are generated by diagrams in which 
virtual and real gauge bosons are radiated by external leg particles, 
and correspond to the soft and collinear singularities appearing in QED 
and QCD, i.e. when massless 
gauge bosons are involved. At variance with this latter case, the weak bosons 
masses put a physical cutoff on these ``singularities", so that virtual 
and real weak bosons corrections can be considered separately. Moreover, 
as the radiation of real weak bosons is in principle detectable, for those 
event selections where one does not include real weak bosons radiation, 
the physical effect of (negative) virtual 
corrections is  singled out, and can amount to tens of per cent. Since these 
corrections 
originate from the infra-red structure of the EW theory, they are  
``process independent" in the sense that they depend only on the external 
on-shell 
legs\cite{Ciafaloni:1998xg,Beccaria:1999fk,Fadin:1999bq,
Ciafaloni:2000df,Ciafaloni:2000rp,Denner:2000jv,Denner:2001gw,
Ciafaloni:2001vt,Beenakker:2001kf,Denner:2003wi,
Kuhn:2004em,Kuhn:2005gv,Kuhn:2007qc,Accomando:2006hq}. 
As shown by Denner and Pozzorini in Refs.~\cite{Denner:2000jv,Denner:2001gw}, 
DL corrections can be accounted for by factorizing a proper correction 
which depends on flavour and kinematics of all possible pairs of electroweak 
charged external legs. SL corrections can be accounted for by factorizing 
an appropriate radiator function associated with each individual external leg. 
Notice that our implementation includes correctly all single logarithmic 
terms of ${\cal O}(\alpha^2 \alpha_s^n)$ of both UV and infrared origin, 
as detailed in Ref.~\cite{Pozzorini:2001rs}. 
The above algorithm has been implemented 
in {\tt ALPGEN v2.1.4}~\cite{Mangano:2002ea}, 
where all the contributing 
tree-level amplitudes are automatically provided. Since the matrix elements in 
{\tt ALPGEN} are calculated within the unitary gauge, for the time being we 
do not implement the corrections for the amplitudes involving longitudinal 
$Z$, which, according to Refs.~\cite{Denner:2000jv,Denner:2001gw}, 
are calculated by means of the Goldstone Boson Equivalence Theorem. 
This approximation affects 
part of the ${\cal O}(\alpha^3)$ and ${\cal O}(\alpha^3 \alpha_s)$ 
contributions, for $Z + 2$~jets and $Z + 3$~jets, respectively, 
and we checked that in view of 
our target precision of few percent it can be 
accepted~\footnote{The logarithms of photonic origin have not been 
considered in this realization since they can be treated separately 
together with their real counterpart 
for the processes under investigation. At any rate, these gauge invariant 
contributions (at the leading order $\alpha_s^{njets} \alpha$) 
for sufficiently inclusive experimental setup give rise to  rather 
moderate corrections. }. 

Despite in this paper we limit 
ourselves to a purely parton-level analysis and a specific signature, 
the implementation 
is completely general. As such it can be generalized to other processes and 
fully matched and showered events can be provided. 

Our numerical results have been obtained by using the code {\tt ALPGEN} 
with default input parameters/PDF set and applying  
two sets of cuts that  
mimic the real experimental event selections of ATLAS and CMS, respectively. 
For $Z + 2$~jets we consider the observable/cuts adopted by ATLAS, namely 
\begin{eqnarray}
&& m_{\rm eff} > 1~{\rm TeV} \qquad \, \, \, \, \, 
\rlap\slash{\!E_T}/m_{\rm eff} > 0.3  \nonumber\\
&& p_T^{j_1} > 130~{\rm GeV} \qquad \, \, 
p_T^{j_2} > 40~{\rm GeV} \quad \, \, |\eta_{j}| < 2.8 \nonumber \\
&& \Delta\phi ({\vec p}_T^j,\rlap\slash{\!\vec{p}_T}) > 0.4 \quad 
\Delta R_{(j_1, j_2)} > 0.4  \, 
\label{eq:atlascut}
\end{eqnarray}
where $j_1$ and $j_2$ are the leading and next-to-leading $p_T$ jets. 
We considered also radiative processes: vector boson pairs plus jets, 
as enumerated in Tab.~\ref{tab:realproc}, in order to give an estimate 
of the (partial) cancellation between virtual NLO and real radiation 
in the presence of a realistic event selection~\footnote{First studies 
on the cancellation between virtual and real EW NLO contributions for 
several processes have been presented 
in Refs.~\cite{Baur:2006sn,Bell:2010gi,Stirling:2012ak}}. 
We consider as real electroweak radiation 
any contribution to 
the experimental event selection of ${\cal O}(\alpha^2 \alpha_s^n)$, 
with $n \leq 2$. In a purely 
perturbative language, only $n=2$ should be considered as ${\cal O}(\alpha)$ 
electroweak corrections (final states in the upper panel of 
Tab.~\ref{tab:realproc}). 
On the other hand, the included additional processes 
contribute to the same experimental signature and moreover 
are the most relevant ones (final states in the lower panel of 
Tab.~\ref{tab:realproc}). 
For the CMS event selection, $n=3$ has to be 
taken instead of $n=2$, as detailed below. In the case of $W$ 
bosons decaying to $l\nu$, 
these contributions are included in the real correction 
only if the charged lepton is lost according to the adopted selection 
criteria. It is worth noting that in our calculation weak bosons are produced 
on shell and decay afterwards. We shall  refer to the coloured partons
present in the matrix element (ME) computation as ME jets and to those 
arising from $V$ 
decay as $V$ jets. In principle, if the ME jets are allowed 
to become unresolved, 
a QCD infrared/collinear problem arises. However, 
in the calculation the ME jets are always required within the acceptance 
cuts, and hence 
no infrared/collinear problem is present. This corresponds to 
a LO prediction of the real 
contribution, that can be considered as a first estimate of the effect. 
The treatment of real QCD radiation with partons below threshold would 
require the inclusion of (not yet available) NNLO QCD corrections 
to $ZV$, for the $Z + 2$~jets signature, and to $ZVj$ for the 
$Z + 3$~jets signature, which could in principle be sizeable, 
but  is beyond the approximation adopted here. 
Moreover, in the presence of the adopted event selections, 
the numerical sizes of $ZVjj(j)$ 
and $ZV(j)$ (the additional jet in parenthesis refers 
to the CMS event selection) are much 
smaller than the dominant $ZVj$, and hence any inaccuracy 
in the estimate of the former 
contributions should be less important at the level of 
the total real radiation 
effect~\footnote{Actually there is a potential ambiguity in disentangling 
EW and
QCD soft/collinear corrections when the ME jets and the $V$ jets overlap. This
contribution is however strongly suppressed by the small available 
phase space and we have checked that it can be estimated to be negligible}. 
In order to give an idea of the hierarchy of the various contributions, 
we report the cross sections, with the event selection 
of Eq.~(\ref{eq:atlascut}), for the three final states 
$ZW(\to \nu {\bar \nu}j j)$, $ZW(\to \nu {\bar \nu} jj) + j$ and 
$ZW(\to \nu {\bar \nu} jj) + jj$~\footnote{Among the processes considered 
in Tab.~\ref{tab:realproc}, these final states are the ones 
with the largest cross sections.} 
at $\sqrt{s} = 14$~TeV (the same hierarchy applies to all other processes 
of Tab.~\ref{tab:realproc}): 
\begin{eqnarray}
\sigma[ZW(\to \nu {\bar \nu}j j)]&=& 0.1911(1)\, {\rm fb}\, , \nonumber \\
\sigma[ZW(\to \nu {\bar \nu} jj) + j]&=&  6.834(1)\, {\rm fb}\, , \nonumber \\
\sigma[ZW(\to \nu {\bar \nu} jj) + jj]&=& 1.213(3)\, {\rm fb} . \nonumber
\end{eqnarray}
\begin {table}
\begin{tabular}{|l|l|l|}
\hline
$ZW(\to \nu_l \bar\nu_l  jj) + jj $           & 
$ZZ(\to \nu_l \bar \nu_l  jj) + jj $          &
$ WW (\to \nu_l l  jj)  +jj $                 \\
\hline
$ ZW (\to \nu_l \nu_l \nu_ll  )  +jj $        &
$ ZW (\to \nu_l lll  )  +jj $                 &
$ ZZ (\to \nu_l \nu_l l l  )  +jj $           \\
\hline
$ ZZ (\to \nu_l \nu_l  \nu_l \nu_l  )  +jj  $ &
$ WW (\to \nu_l \nu_l l l  )  +jj  $          &
$ZW(\to \nu_l l  jj)+jj $                                  \\
\hline
\hline
$ZW(\to \nu_l \bar \nu_l jj) $      & 
$ZW(\to \nu_l l  jj) $              & 
$ZZ(\to \nu_l \bar \nu_l  jj)$      \\
\hline
$ WW (\to \nu_l l  jj)$             &
$ZW(\to \nu_l l  jj)+j $            &
$ZW(\to \nu_l \bar\nu_l  jj) + j $  \\
\hline
$ZZ(\to \nu_l \bar \nu_l  jj) + j $ &
$ WW (\to \nu_l l  jj)  +j $        & 
                                   \\
\hline
\end{tabular}
\caption{\label{tab:realproc}
Vector boson radiation processes contributing to the considered signatures. 
In parenthesis we specify vector boson decay channels, while outside 
the parenthesis 
$j$ stands for a matrix element QCD parton. The above processes are for the 
$Z + 2$~jet final state, whereas for three jet final states the processes 
are the same ones plus an additional QCD parton.}
\end{table}
The definition of the event selection for real radiation processes 
requires further details on the adopted event selection w.r.t. 
Eq.~(\ref{eq:atlascut}), 
in order to mimic, in a simplified way, the ATLAS procedure. 
Missing transverse energy is defined as 
$\vec{\rlap\slash{\!H_T}} = - \sum_i \vec{p_t}_i$, 
where $i$ is either a 
tagged jet or a jet with $p_{Tj}< 40$~GeV or $2.8 < |\eta_j| < 4.5$ 
(in our simulation this is necessarily a jet coming from vector boson decay), 
or an untagged charged lepton. By tagged jet we mean a jet 
with $|\eta_j| < 2.8$ and $p_{Tj}> 40$ GeV. Jets from vector boson 
decays are recombined with other jets if they fall within a separation 
cone with radius $R = 0.4$. The event is discarded if it contains a tagged 
charged lepton, i.e. a lepton ($e$, $\mu$ or $\tau$) with 
$p_T > 20$~GeV and $\vert \eta \vert < 2.4$. For the tagged jets, 
an additional requirement is imposed: if $\Delta R_{jl} > 0.2$, 
the jet is considered untagged. After this step, the leptons 
with a separation from any tagged jet $\Delta R_{jl}< 0.4$ are considered 
untagged. Finally the event is accepted if it contains exactly two tagged jets 
and no surviving tagged lepton and it satisfies the cuts 
of Eq.~(\ref{eq:atlascut}). 

For the $Z+3$~jets final state we consider the observables/cuts used by 
CMS~\footnote{For the sake of simplicity, we consider the exclusive signature 
$Z + 3$~jets instead of the inclusive one, as done by CMS.}, namely
\begin{eqnarray}
&& H_T > 500~{\rm GeV} \qquad \, \, \, \, \, \, 
|\rlap\slash{\!\vec{H}_T}| > 200~{\rm GeV}  \nonumber\\
&& p_T^j > 50~{\rm GeV} \qquad  \, \, |\eta_j| < 2.5 
\quad \Delta R_{(j_i, j_k)} > 0.5 \nonumber\\
&& \Delta\phi ({\vec p}_T^{j_1, j_2},\rlap\slash{\!{\vec H}_T}) > 0.5 
\qquad  \Delta\phi (\vec{p}_T^{j_3},\rlap\slash{\!{\vec H}_T}) > 0.3 \nonumber
\end{eqnarray}
Concerning additional real vector boson radiation, 
in this case the missing transverse energy receives contribution from 
tagged jets only, namely jets with $p_{T}^j> 50$~GeV and 
$|\eta_j| < 2.5$. Jets from vector boson 
decays are recombined with other jets if they fall within a separation 
cone with $R = 0.5$ and charged leptons with $\Delta R_{jl}< 0.2$ are 
recombined as well. Events with tagged surviving (not recombined with jets) 
charged leptons are discarded. Leptons are tagged if $p_{T}^l > 10$~GeV 
and $|\eta_l| < 2.5$. 

\begin{figure}[h]
\includegraphics[scale=0.45]{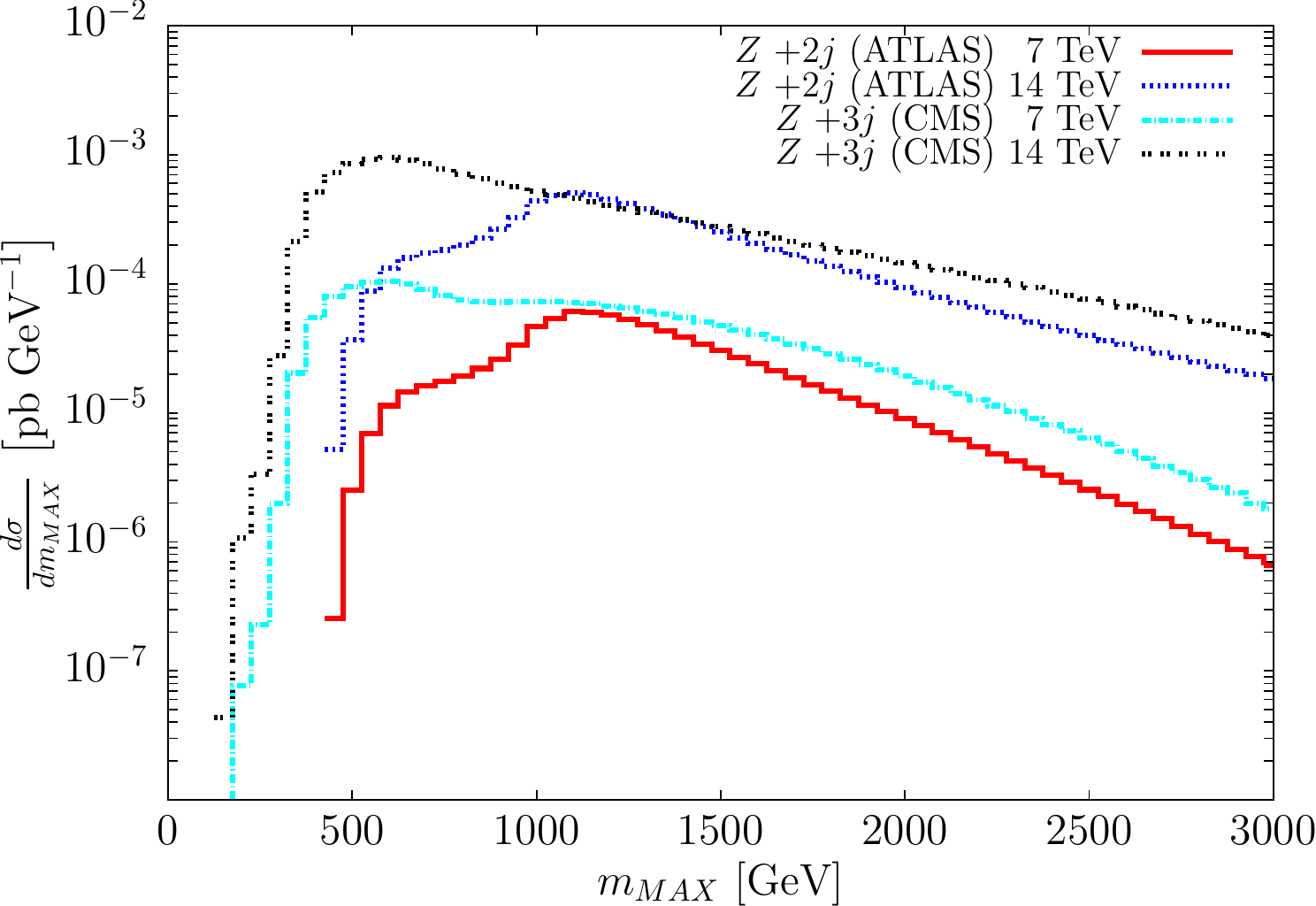}
\caption{\label{fig:minvpairs} $Z+2,3$~jets: distributions of the maximum 
invariant mass at $\sqrt{s} = 7,14$~TeV}
\end{figure}

As a first test, it is worth assessing the applicability of the theoretical 
 approach described above for the virtual EW Sudakov corrections.  
In Refs.~\cite{Denner:2000jv,Denner:2001gw} the underlying hypothesis is that 
all kinematical invariants are much larger than $M_W$. 
Fig.~\ref{fig:minvpairs} shows the maximum invariant 
mass distributions for the processes $Z + 2,3$~jets at $\sqrt{s} = 7,14$~TeV, 
obtained by considering, on an event by event basis, all possible combinations 
of invariant masses between electroweak charged particles at the parton level. 
One can notice that most of the 
events are characterized by at least one invariant mass above, say, 500~GeV. 
We expect that the 
approximation of Refs.~\cite{Denner:2000jv,Denner:2001gw} 
still holds, since radiator contributions depending on 
large kinematical invariants are reliable, whereas those depending on 
small kinematical invariants (at any rate of the order of $M_W$, 
as ensured by the applied cuts) lead to unreliable contributions, which, 
however, are numerically below the stated accuracy, since the involved 
logs are of order one or below. The above argument has been validated with 
results available in the literature as follows: first, we compared the 
predictions for $Z + 1$~jet and $W + 1$~jet with 
Refs.~\cite{Kuhn:2004em,Kuhn:2007qc}, finding 
a level of agreement better than 1\%; second, we estimated the corrections to 
$p_T^Z$ and to the leading jet $p_T$ distributions in the large tails 
for the process $Z + 2 $~jets with only one fermionic current, 
as discussed in Ref.~{\cite{Actis:2012qn}}, finding good agreement. For the 
same kind of process we cross-checked our results with the automatic package 
{\tt GOSAM v1.0}~\cite{Cullen:2011ac}, with the event selection adopted in 
the present study. Since the electroweak renormalization is not yet available 
in the present version of {\tt GOSAM}, we subtracted the logarithmic terms due 
to the renormalization counterterms from the formulae of 
Refs.~\cite{Denner:2000jv,Denner:2001gw} and tested the asymptotic behaviour 
of all relevant distributions. In particular we performed this analysis for 
different subprocesses: $q \bar q \to Z g g$, $q \bar q \to Z q'' \bar q''$, 
$q q \to Z q q$ and $q q' \to Z q q'$ (with $q$ and $q'$ belonging to the same 
isodoublet). For all the above cases we found that the shape of the 
distributions 
predicted by the two calculations is in good agreement. 
In particular, the relative weight 
of two-quark and four-quark subprocesses is about 75\% and 25\% 
for total cross sections, while 
for the observables under consideration and in the high tailes 
is about 50\% each at the LO, respectively. 

\begin{figure}[h]
\includegraphics[scale=0.43]{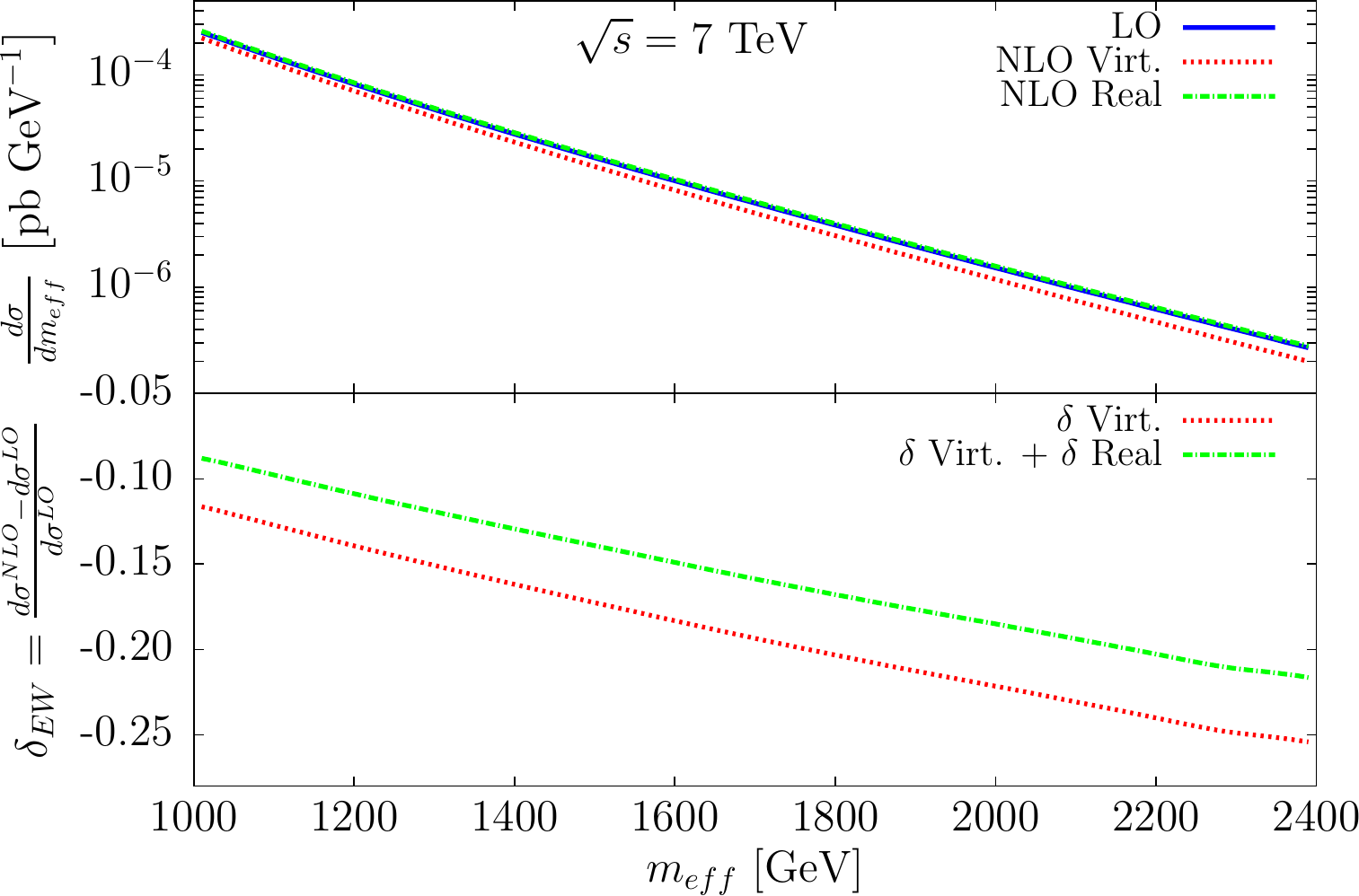} \\
\includegraphics[scale=0.43]{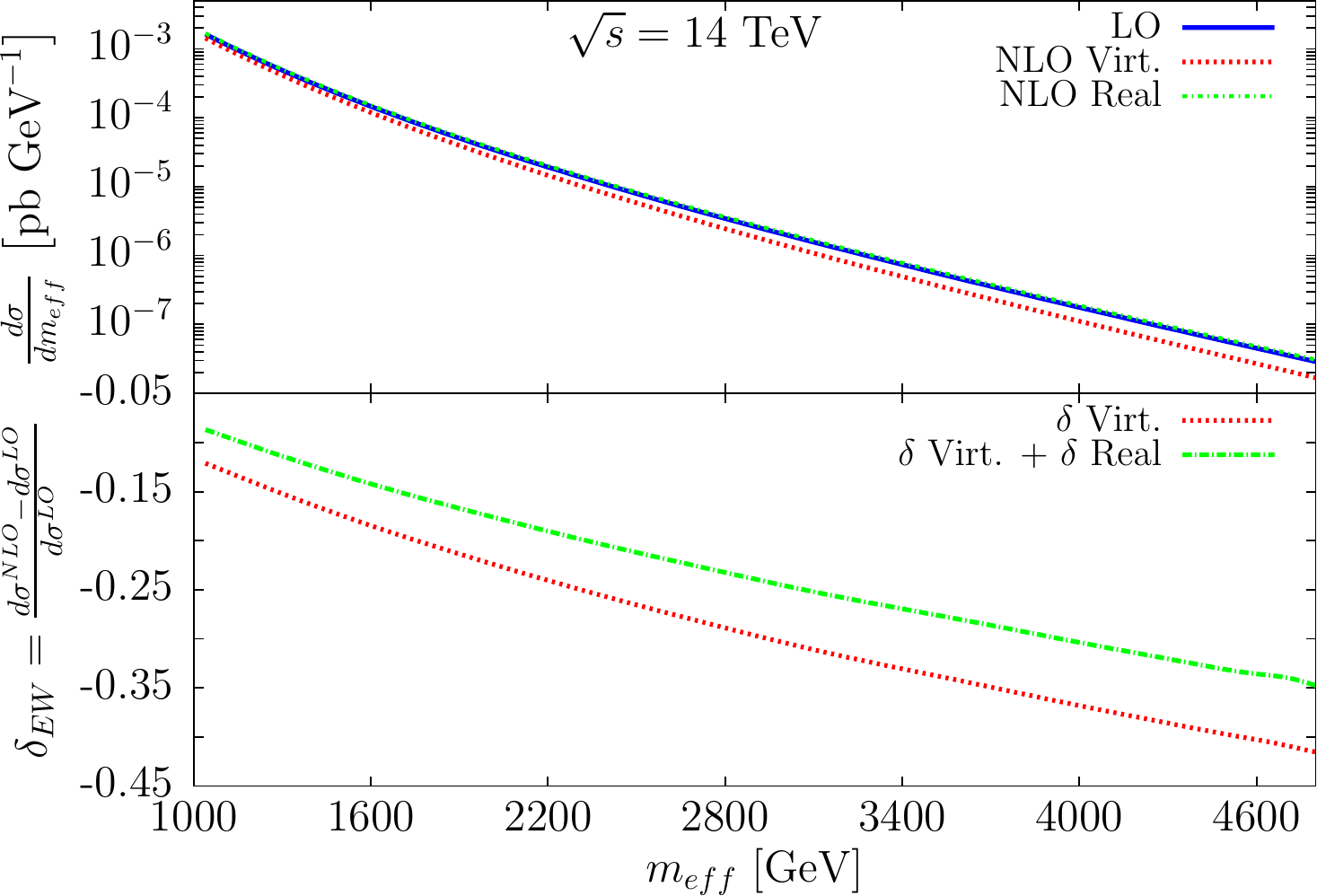}
\caption{\label{fig:meff2j} $Z+2$~jets: ATLAS $m_{\rm eff}$ and EW correction 
at $\sqrt{s} = 7, 14$~TeV}
\end{figure}

Fig.~\ref{fig:meff2j} shows the effect of the Sudakov logs 
on the effective mass 
distribution in the process $Z+2$~jets under ATLAS conditions. 
In both windows, the upper panel displays the effective mass 
distribution at LO 
(blue, solid) and including the approximate NLO virtual corrections 
(red, dotted) due to weak bosons in the 
Sudakov limit as given by Denner and Pozzorini, respectively. 
The effect of the inclusion of real radiation 
of electroweak bosons is also shown (green, dash-dotted). 
The lower panel represents the relative 
correction due to virtual contributions only (red, dotted) 
and the sum of virtual and real 
radiation (green, dash-dotted). The two windows correspond 
to $\sqrt{s} = 7,14$~TeV, respectively. 
As can be seen, the negative correction due to Sudakov logs is of the order of 
some tens of per cent, raising to about 40\% in the extreme region at 
$\sqrt{s} = 14$~TeV. Real radiation partially cancels the effect, 
introducing a positive contribution of about some per cent. 
\begin{figure}[h]
\includegraphics[scale=0.43]{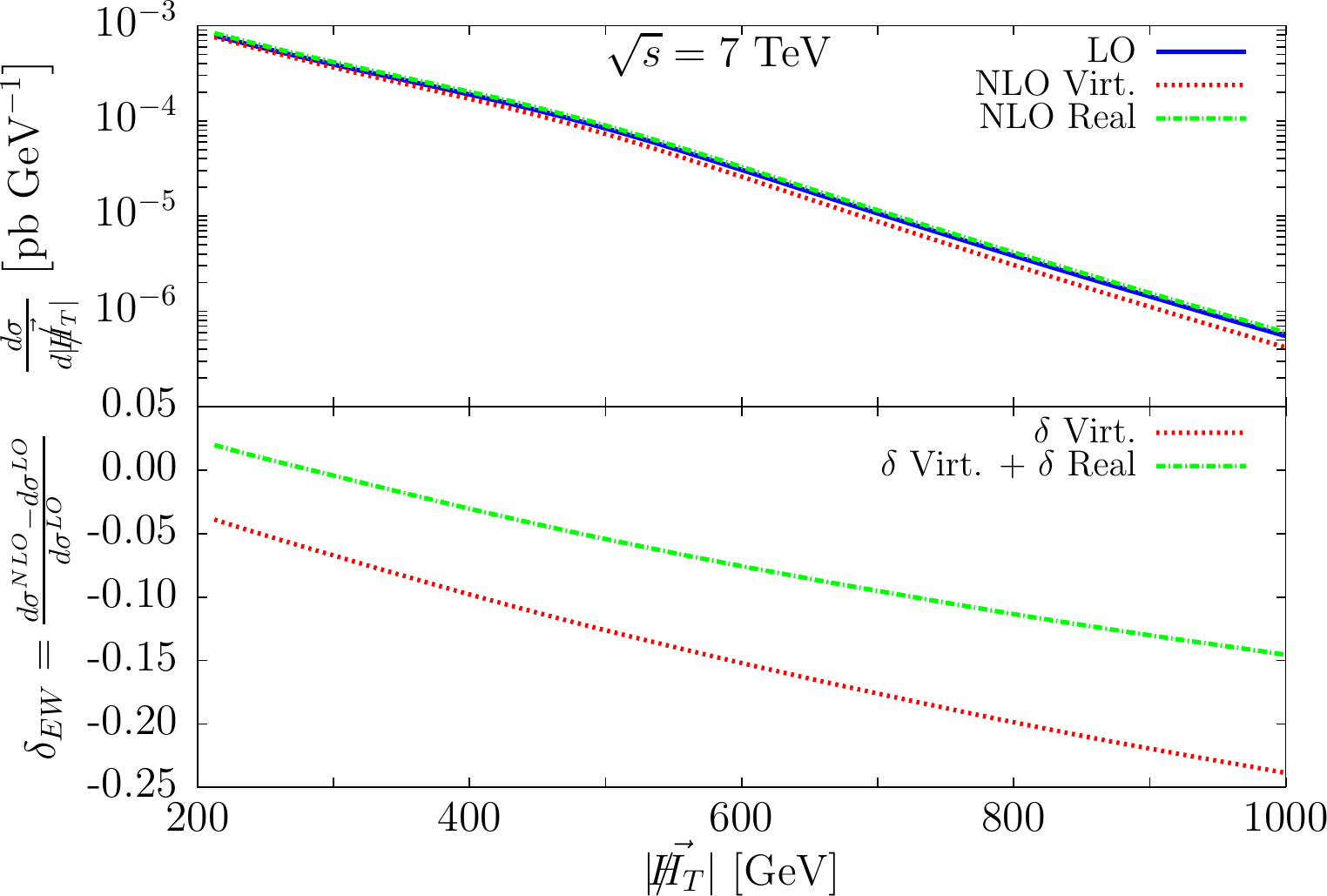} \\
\includegraphics[scale=0.43]{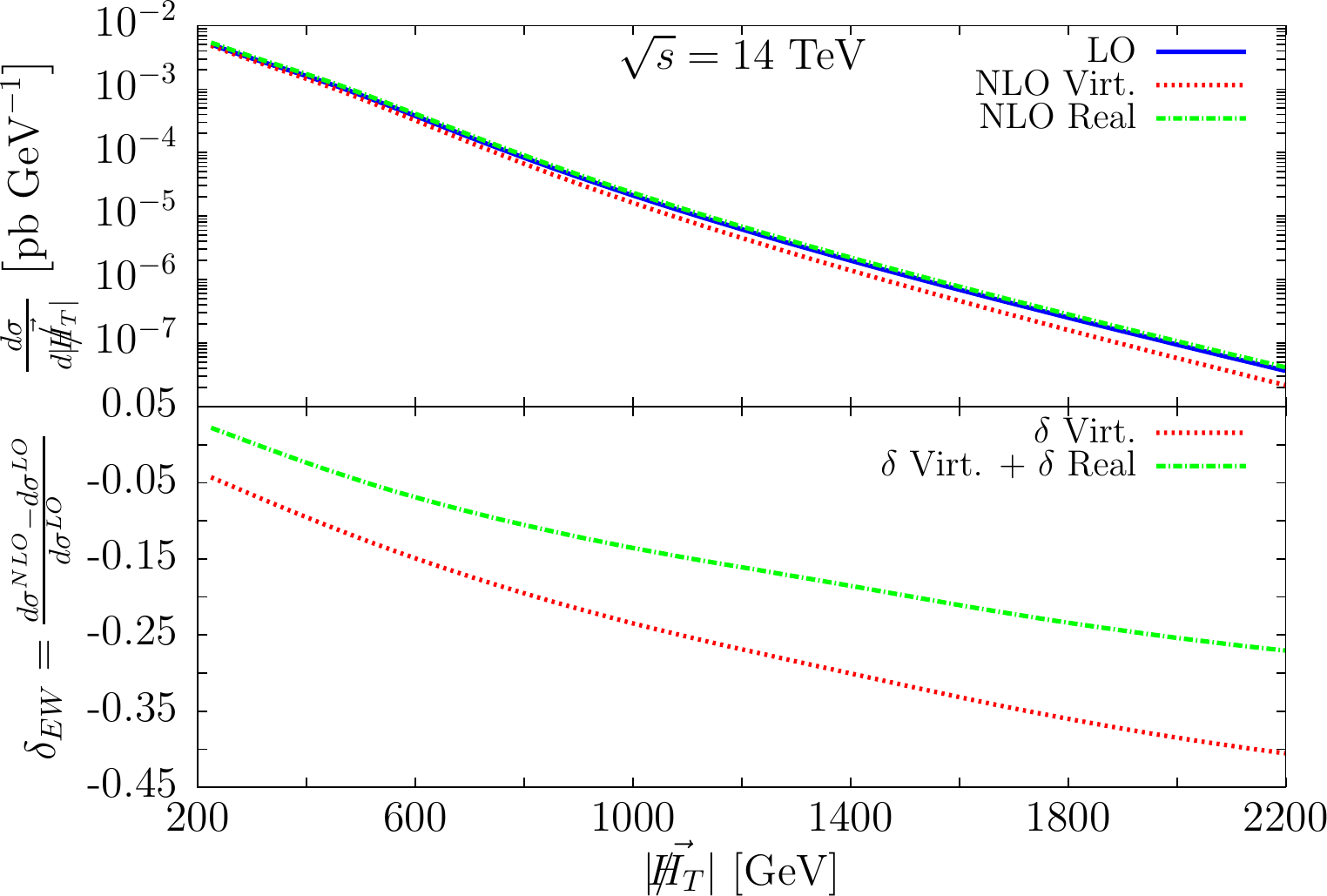}%
\caption{\label{fig:Z3jEWcorr} $Z+3$~jets: CMS $|\rlap\slash{\!H_T}|$ 
and EW correction at $\sqrt{s} = 7, 14$~TeV}
\end{figure}
Fig.~\ref{fig:Z3jEWcorr} is the analogous of Fig.~\ref{fig:meff2j} 
for the observable $|\rlap\slash{\!{\vec H}_T}|$ under CMS conditions. 
Also in this case the virtual correction is large and negative, 
reaching the value of -25\% at $\sqrt{s} = 7$~TeV and -45\% at 
$\sqrt{s} = 14$~TeV. 
In these experimental conditions real radiation gives a positive 
correction slightly 
larger than before, that can be as large as about +15\%. 

To summarize, we computed the NLO  Sudakov EW corrections to $Z + n$~jets, 
$n = 1,2,3$, as the main background to 
NP searches at the LHC. We found that such corrections represent 
a sizeable effect, of the order of 
tens of per cents, that has to be taken into account,  together with 
the partially compensating 
contribution of weak bosons real radiation. 
The described calculation represents the first implementation 
of the Denner-Pozzorini algorithm 
in a multiparton LO generator, and paves the way to future applications 
to other multijet 
production processes at the energy frontier.

\begin{acknowledgments}
This work was supported in part by the Research Executive Agency (REA) of the 
European Union under the Grant Agreement number 
PITN-GA-2010-264564 (LHCPhenoNet), and by the Italian Ministry of University 
and Research under the 
PRIN project 2010YJ2NYW.
The work of L.B. is supported by the 
ERC grant 291377, ``LHCtheory - 
Theoretical predictions and analyses of LHC physics: advancing the precision 
frontier". F.P. would like to thank the CERN PH-TH Department 
for partial support and 
hospitality during several stages of the work. 
\end{acknowledgments}

\appendix

\bibliography{Sudakov}

\end{document}